\documentclass{jaa}
\usepackage[square]{natbib}
\usepackage[colorlinks=true, citecolor = blue]{hyperref}
\newcommand{\hi}{H{\sc i}}

\usepackage{graphicx}
\usepackage{amsmath}

\begin{document}\sloppy

\title{Exploring Machine Learning Regression Models for Advancing Foreground Mitigation and Global 21cm Signal Parameter Extraction}


\author{Anshuman Tripathi\textsuperscript{1*}, Abhirup Datta\textsuperscript{1}, and Gursharanjit Kaur\textsuperscript{2}}  



\affilOne{\textsuperscript{1}Department of Astronomy, Astrophysics, and Space Engineering, Indian Institute of Technology Indore, Madhya Pradesh, 453552, India.\\}
\affilTwo{\textsuperscript{2}Center for Theoretical Physics, Polish Academy of Sciences, al. Lotnikow 32/46, 02-668 Warsaw, Poland\\}

\date {}
\twocolumn[{

\maketitle

\corres{anshumantripathi85@gmail.com}

\msinfo{}{}

\begin{abstract}
Extracting parameters from the global 21cm signal is crucial for understanding the early Universe. However, detecting the 21cm signal is challenging due to the brighter foreground and associated observational difficulties. In this study, we evaluate the performance of various machine-learning regression models to improve parameter extraction and foreground removal. This evaluation is essential for selecting the most suitable machine learning regression model based on computational efficiency and predictive accuracy. We compare four models: Random Forest Regressor (RFR), Gaussian Process Regressor (GPR), Support Vector Regressor (SVR), and Artificial Neural Networks (ANN). The comparison is based on metrics such as the root mean square error (RMSE) and \( \rm R^{2} \) scores. We examine their effectiveness across different dataset sizes and conditions, including scenarios with foreground contamination. Our results indicate that ANN consistently outperforms the other models, achieving the lowest RMSE and the highest \( \rm R^{2} \) scores across multiple cases. While GPR also performs well, it is computationally intensive, requiring significant RAM and longer execution times. SVR struggles with large datasets due to its high computational costs, and RFR demonstrates the weakest accuracy among the models tested. We also found that employing Principal Component Analysis (PCA) as a preprocessing step significantly enhances model performance, especially in the presence of foregrounds. 


\end{abstract}


}]



\doinum{12.3456/s78910-011-012-3}
\artcitid{\#\#\#\#}
\volnum{000}
\year{0000}
\pgrange{1--}
\setcounter{page}{1}
\lp{1}

\section{Introduction}

21cm cosmology is a key scientific objective for upcoming experiments and is emerging as a powerful tool for probing the Universe across extensive cosmic timescales. This approach leverages the 21cm hyperfine transition of neutral hydrogen (HI), caused by the spin-flip transition in hydrogen atoms \citep{Field1958, 1959ApJ...129..536F}, to probe critical epochs in cosmic history, including the Dark Ages, the Cosmic Dawn, and the Epoch of Reionization (EoR) \citep{2006Furlanetto_a, 2006Fan, pritchard201221}. These periods are pivotal for understanding the formation of the first stars, galaxies, and black holes and the transition of the intergalactic medium (IGM) from a neutral state to an ionized state \cite{2010Morales, 2016Barkana, 2018Dayal}. In the $\Lambda$CDM cosmological model, the first luminous structures formed when hydrogen collapsed under gravity, this era is known as cosmic dawn. The ionization of the IGM was initiated by ultraviolet (UV) photons emitted by these early sources, marking the onset of reionization during the EoR. However, this epoch remains poorly constrained due to the lack of direct observational probes at such high redshifts. The HI 21cm signal stands out as a promising probe for investigating these unexplored redshift regimes \citep{2013ASSL..396...45Z}. Unlike the Lyman-$\alpha$ signal, the 21cm signal does not saturate at high optical depths, allowing it to trace the evolution of cosmic structures and the IGM across different redshifts \citep{1959ApJ...129..536F, Field1958, 1990Scott}. By capturing the properties and evolution of the early Universe, the 21cm signal offers a unique window into cosmic history, making it a vital focus for advancing our understanding of the cosmos.

Detecting a 21cm signal presents several challenges. The signal is overwhelmed by bright galactic \citep{1999Shaver} and extragalactic foregrounds \citep{2002Di_Matteo, 2004Di_matteo}, which are $10^{4}$ to $10^{5}$ times more bright within the redshifted frequency range. Additionally, ionospheric effects and the instrument's response to the observations further complicate achieving accurate measurements \citep{2014Vedantham, 2016Datta, 2021Shen}. Two separate experimental methodologies are used to observe these signals. One method utilizes single-radiometer antennas, as seen in experiments as the Experiment to Detect the Global Epoch of Reionization Signature (EDGES) \cite{2018Natur.555...67B}, Shaped Antenna Measurement of the Background Radio Spectrum (SARAS) \cite{2022NatAs.tmp...47S}, and Large-Aperture Experiment to Detect the Dark Ages (LEDA) \cite{leda}. Alternatively, interferometers such as the Hydrogen Epoch of Reionization Array (HERA), Low-Frequency Array (LOFAR) \cite{lofar2013}, and the forthcoming Square Kilometer Array (SKA) \cite{ska2015} are used. The EDGES experiment has reported a potential detection of the sky-averaged global 21cm signal, featuring an absorption trough centred at 78 MHz with an anomalously large depth—approximately twice that predicted by standard cosmological models \citep{2018Natur.555...67B}. If confirmed, this result could offer profound insights into the physics of the cosmic dawn and the epoch of reionization. However, independent analysis from the SARAS experiment \cite{2022NatAs.tmp...47S} rejected the EDGES signal profile with 95.3 \% confidence, instead finding results consistent with the predictions of standard cosmology. This discrepancy highlights the importance of accurately modelling and removing distortions from the observed signal to ensure robust and reliable interpretation. Consequently, it is essential to comprehend and eliminate any contaminating variables from the investigation.

Machine learning (ML) methods have been widely used for parameter estimation and signal modelling in the last several years in the various fields of cosmology. To fully emulate the 21 cm signal, several methods have been proposed, including those based on artificial neural networks (ANNs) \citep{jennings2019evaluating, cohen2020emulating, schmit2018emulation, globalemu, VAE, tiwari2022, 2025Tripathi} and long short-term memory networks (LSTMs) \citep{2024LSTM}. In addition to signal modelling, these machine learning algorithms have been employed to extract signal and related parameters \citep{choudhury2020extracting, choudhury2021using, Tripathi2024, Tripathi2024_Samp}.
 Alongside machine learning techniques, several conventional approaches are also used for global 21cm signal and parameter extraction. These include Markov Chain Monte Carlo (MCMC) for posterior exploration \cite{harker2012mcmc, Rapetti_2020}, Singular Value Decomposition (SVD) for constructing foreground and signal models \cite{2018Tauscher, Tauscher2021}, Maximally Smooth Functions (MSFs) for constrained fitting of smooth foregrounds~\cite{2015Mayuri, Bevins_2021}, a Bayesian nested sampling algorithm based on parametrized, physical-motivated modelling of the foregrounds presented \cite{Anstey2021} and the Vari-Zeroth-Order Polynomial (VZOP) for improved polynomial fitting techniques \citep{Liu2024}.

In the rapidly evolving field of machine learning, selecting the appropriate model that effectively balances predictive accuracy and computational cost remains a critical challenge, especially in applications such as {\hi} 21 cm signal analysis. In this study, we systematically evaluate four widely used machine-learning regression techniques: Gaussian Process Regression (GPR), Random Forest Regression (RFR), Support Vector Regression (SVR), and Artificial Neural Networks (ANNs) with the goal of improving foreground removal and parameter extraction for the global 21cm signal.

Our benchmarking is conducted under two scenarios: (i) an idealized case with only the 21cm signal, and (ii) a more realistic case that includes both foregrounds and thermal noise. For each model, we assess prediction accuracy, computational time, and memory usage using the same hardware configuration (\texttt{Intel(R) Xeon(R) Gold 5320 CPU @ 2.20GHz} and 755 GB of RAM). We also explore how the performance of each model scales with the size of the training dataset, to identify models that remain effective even with limited data availability.

Among these machine learning methods, ANNs and GPR have already been widely applied in 21\,cm cosmology, particularly for parameter extraction, signal emulation, and foreground subtraction. By contrast, RFR and SVR have not yet been explored in 21\,cm applications, though they have been successfully employed to tackle other cosmological problems and parameter estimation tasks. By systematically comparing these established models (ANNs, GPR) alongside RFR and SVR within a unified framework, our analysis provides a comprehensive assessment of their strengths and limitations.  

ANNs have proven particularly valuable in global-signal experiments, where they have been used to extract {\hi} signal parameters while mitigating the effects of foreground contamination, ionospheric distortions, and instrumental systematics \citep{choudhury2020extracting, choudhury2021using, Tripathi2024, Tripathi2024_Samp}. GPR, on the other hand, has been predominantly applied in the context of interferometric observations. It was first introduced by the LOFAR collaboration for foreground subtraction \citep{Mertens2018} and subsequently used for setting upper limits on the 21\,cm power spectrum \citep{Gehlot2019, Mertens2020}. Beyond this, GPR has demonstrated exceptional efficiency as an emulator in 21\,cm cosmology. For instance, \citet{2025Pundir} developed a GPR-based emulator trained on low-dynamic-range N-body simulations to predict {\hi} density fields during the Epoch of Reionization, achieving better than 10\% accuracy in large-scale power spectra. Similarly, \citet{2024Chouhury} introduced a GPR-based likelihood emulator for the semi-numerical reionization code \texttt{SCRIPT}, enabling rapid parameter inference from mock 21\,cm data with nearly an order-of-magnitude reduction in computational time. Earlier, \citet{2023Maity} proposed a GPR-trained \texttt{SCRIPT} framework to efficiently explore reionization parameter space, significantly reducing MCMC runtimes.  

RFR has been shown to extract cosmological parameters from the matter power spectrum with good accuracy for $\Omega_m$ and $\sigma_8$, though its performance degrades for parameters such as $h$ and $n_s$, where neural networks yield superior results \citep{2021Lazanu}. Likewise, Support Vector Machines (SVMs) have been applied to address the $H_0$ tension. In tests using synthetic expansion-rate data, SVMs demonstrated the best bias--variance trade-off among several regression methods, providing a competitive alternative to Gaussian Processes \citep{2023Bengaly}.

This work offers practical insights into selecting optimal regression strategies that balance computational efficiency and predictive performance for present and future global 21cm signal studies.

This paper is organized as follows: Section \ref{21cm_sig} discusses the observable features of the Global 21cm Signal. Section \ref{obs_challenge} addresses the key observational challenges. Section \ref{method} details the methodology used for simulating both the global 21cm signal and foregrounds. Section \ref{ml_models} provides an overview of the machine learning regression models employed. Section \ref{ttd} describes the preparation of datasets for training and testing. Section \ref{result} presents the results, including a comparative analysis of the performance of all four ML models. Finally, Section \ref{con} offers a summary and concluding discussion.

\section{Observable Global 21cm Signal}\label{21cm_sig}
The hyperfine splitting of the 1S ground state of the hydrogen atom arises from the interaction between the magnetic moments of the proton and the electron. This interaction produces the notable 21 cm signal. The process is commonly known as the spin-flip transition, wherein the spins shift from a parallel to an anti-parallel alignment. This transition results in the spontaneous emission of a photon with a wavelength of 21 cm.

In a single-radiometer experiment, the differential brightness temperature is determined by comparing the signal's brightness temperature $T_{b}$ to the background temperature $T_{CMB}$, which is the Cosmic Microwave Background (CMB) \citep{2006Furlanetto, mirocha2014decoding}.
\begin{align}
\delta T_{b} \approx 27(1 - x_{HI})\left (\frac{\Omega _{b}h^2}{0.023} \right) \left(\frac{0.15}{\Omega _{m,0}}\frac{1+z}{10}\right) ^{1/2} \left(1 - \frac{T_{CMB}(z)}{T_{s}}\right) \end{align}
where $\Omega_{b}$ represents the baryon density in units of the critical density, $\Omega_{m}$ the total matter density, $H(z)$ the Hubble parameter at redshift $z$, and $T_{s}$ the spin temperature of neutral hydrogen. In this context, $x_{HI}$ is the neutral percentage of hydrogen.

The spin temperature ($\rm T_{s}$) of hydrogen atoms, which determines the relative populations in the two spin states, is regulated by the interplay of three processes: absorption and stimulated emission of CMB photons, controlled by the CMB temperature ($\rm T_{CMB}$); collisions with hydrogen atoms, free electrons, and protons, dictated by the kinetic gas temperature ($\rm T_{k}$) of the intergalactic medium (IGM); and scattering of Lyman-$\alpha$ photons, defined by the color temperature ($\rm T_{\alpha}$) in the Wouthuysen–Field effect \cite{1952Wouthuysen}. The spin temperature, $\rm T_{s}$, is determined by following \citep{1959ApJ...129..536F, 10.1088/2514-3433/ab4a73ch1}:

\begin{align}
T_{s}^{-1} = \frac{T_{CMB}^{-1} + x_{k} T_{k}^{-1} + x_{\alpha} T_{\alpha}^{-1} }{1 + x_{k} + x_{\alpha}}
\label{eq:couple}
\end{align} 
Here, $x_{k}, x_{\alpha}$ are collisional and Lyman- $\alpha$ coupling coefficients. Thus, the global signal evolves over the redshift range as a function of the properties of IGM.

\section{Observational Challenges}\label{obs_challenge}
Detecting the redshifted 21 cm signal faces several observational challenges, including bright foreground emissions, ionospheric disturbances, beam chromaticity, thermal noise, and radio frequency interference (RFI). This study primarily focuses on addressing two main challenges: foreground contamination and thermal noise. To do this, we simulate observations to create training datasets.

\subsection{Foregrounds}
Foreground poses a significant observational challenge for detecting the 21cm signal, as it is dominated by galactic synchrotron emission, free-free radiation, and thermal dust, which are several orders of magnitude brighter than the signal \cite{1999Shaver, 2003Oh}. Extragalactic sources, such as radio emissions from star-forming galaxies, further complicate the measurement \cite{2002Di_Matteo, 2004Di_matteo}. The spectral smoothness of these foregrounds compared to the 21cm signal allows for modelling and subtraction, but inaccuracies in foreground removal can introduce residuals that obscure the cosmological information. The high-frequency coherence of galactic foregrounds facilitates subtraction \citep{liu2009improved}, and their smooth spectral nature allows for representation using low-order polynomials \citep{pritchard201221, 2015Bernardi}.

\subsection{Thermal Noise}
The thermal noise in the observed spectrum, represented as $n(\nu)$, can be described using the ideal radiometer equation as follows:
\begin{equation}
\begin{split}
{n(\nu) \approx \frac{T_{sys}(\nu)}{\sqrt{\delta \nu  \cdot \tau}}},
\end{split}
\end{equation}


In this context, $\rm T_{sys}(\nu)$ represents the system temperature, $\delta \nu$ is the observational bandwidth, and $\tau$ denotes the observation time.

\section{Methodology}\label{method}
\subsection{Simulation of Global 21cm} \label{ares}
To simulate the global 21cm signal from CD and EoR,  we have used the tanh parameterized model based on ARES parameterization \citep{2012Mirocha, 2014MNRAS.443.1211M}.  The model adopts simple parametric representations for the Lyman-$\alpha$ background, IGM temperature, and reionization histories \citep{2016MNRAS.455.3829H}. The parameterized tanh-based model, in particular, serves as an intermediate approach between purely phenomenological models (e.g., turning points) and fully physical simulations \citep{mirocha2015interpreting}, it is computationally efficient while maintaining a direct connection to the thermal and ionization state of the IGM. It characterizes the signal based on the properties of IGM, such as the Lyman-$\alpha$ coupling strength, $J_{\alpha}(z)$; the temperature of the IGM, $T(z)$; and the ionization fraction, ${\overline{X}_{i}}$, without incorporating the details of the source properties. The evolution of these parameters is described as follows:

\begin{equation}
 A(z) = \frac{A_{ref}}{2}\left(1 + tanh\left(\frac{z_{0} - z}{\Delta z}\right)\right)
 \label{tanh_equation}
\end{equation}

where, $ \rm A_{ref}$ represents the step height, $z_{0}$ denotes the pivot redshift, and $\Delta z$ specifies the width. These quantities are zero at high redshifts, become active within a redshift interval $\Delta z$ centered around $z_{0}$, and reach maximum saturation at $A_{ref}$ at lower redshifts.

The step height, \( A_{\text{ref}} \), associated with the Lyman-\(\alpha\) (\(Ly\alpha\)) background, is represented by \( J_{\text{ref}} \) and saturates at low redshift with a value of 11.69, as determined by \cite{2016MNRAS.455.3829H} using MCMC parameter estimation. In this study, it is treated as a constant. The redshift interval and pivot redshift for the \(Ly\alpha\) background tanh parameterization are denoted by \( J_{dz} \) and \( J_{z0} \), respectively. For X-ray heating, the intergalactic medium (IGM) temperature, \( T(z) \), is characterized by \( T_{dz} \) and \( T_{z0} \), which represent the redshift interval and central redshift in Kelvin. The corresponding amplitude, \( A_{\text{ref}} \), associated with \( T_{\text{ref}} \), saturates at around 1000 K. Regarding ionization fraction (\(\overline{X}_{i}\)), the natural value for the step height is unity. The ionization process occurs over a redshift interval \( X_{dz} \) and pivot redshift \( X_{z0} \).  

The parameter values inferred by \cite{2016MNRAS.455.3829H} are:  
\begin{itemize}
    \item \( J_{dz} = 3.31 \), \( J_{z0} = 18.54 \) (Lyman-\(\alpha\) background)  
    \item \( T_{dz} = 2.82 \), \( T_{z0} = 9.77 \) (IGM temperature)  
    \item \( X_{dz} = 2.83 \), \( X_{z0} = 8.68 \) (Ionization fraction)  
\end{itemize}

\subsection{Foreground Simulation}
The foregrounds dominate global 21cm observations, surpassing the signal by several orders of magnitude. Simulating these foregrounds is crucial for mitigation strategies and detecting the 21cm signal. Major components include Galactic synchrotron and free-free emission, extragalactic point sources, and the ionosphere. Modeling astrophysical foregrounds is essential for isolating the faint cosmological signal. In this work, we modeled the diffuse foreground as a third-order polynomial in $log(\nu)-log(T)$ space, consistent with the approaches in \citep{choudhury2020extracting, harker2015selection}.

\begin{align}
\log(T_{FG}) =  \sum_{i = 0}^{n}a_{i}\left(\log \left(\dfrac{\nu}{\nu_{0}}\right)\right)
\end{align}

The foreground model, defined at a reference frequency of \( \nu_0 = 80 \, \text{MHz} \), uses a logarithmic polynomial to relate frequency (\( \nu \)) and temperature (\( T \)).

\section{Overview of the ML models}\label{ml_models} 
In this study, we compare the performance of various machine learning (ML) regression models in extracting global 21cm signal parameters. This analysis considers two scenarios: one where only the signal is present and another where foreground and thermal noise overshadow it. Additionally, we evaluate the computational resources required for successfully running these ML algorithms to recover both signal and foreground parameters in each scenario. A detailed overview of each ML regression model is provided below.

\subsection{Artificial Neural Network (ANN)}
ANN are computational models inspired by the structure and function of the human brain, designed to recognize patterns and relationships in data. They consist of layers of interconnected nodes, or "virtual neurons," that process information to generate outputs. ANNs are widely applied in tasks such as image recognition, natural language processing, and prediction, owing to their capacity to learn and handle complex data \citep{zupan1994introduction, agatonovic2000basic}.

ANNs consists of three fundamental layers: the input layer, one or more hidden layers, and the output layer. The depth of the network is determined by the number of hidden layers, while its width is defined by the number of neurons per layer. In feed-forward architectures, information moves in a single direction, with each neuron in one layer fully connected to those in the next. These connections are governed by weights ($w_{ij}$) and biases ($b_j$), with $x_j$ representing the input data to the $j$th neuron.

Analytically, this can also be represented as \citep{Tripathi2024},
\begin{equation}
    Y_i = \sum_{j=1}^{L} w_{ij}x_j + b_j
\end{equation}

\begin{figure}[ht]
    \centering
    \includegraphics[width=0.49\textwidth, trim=5cm 5cm 1cm 11cm, clip]{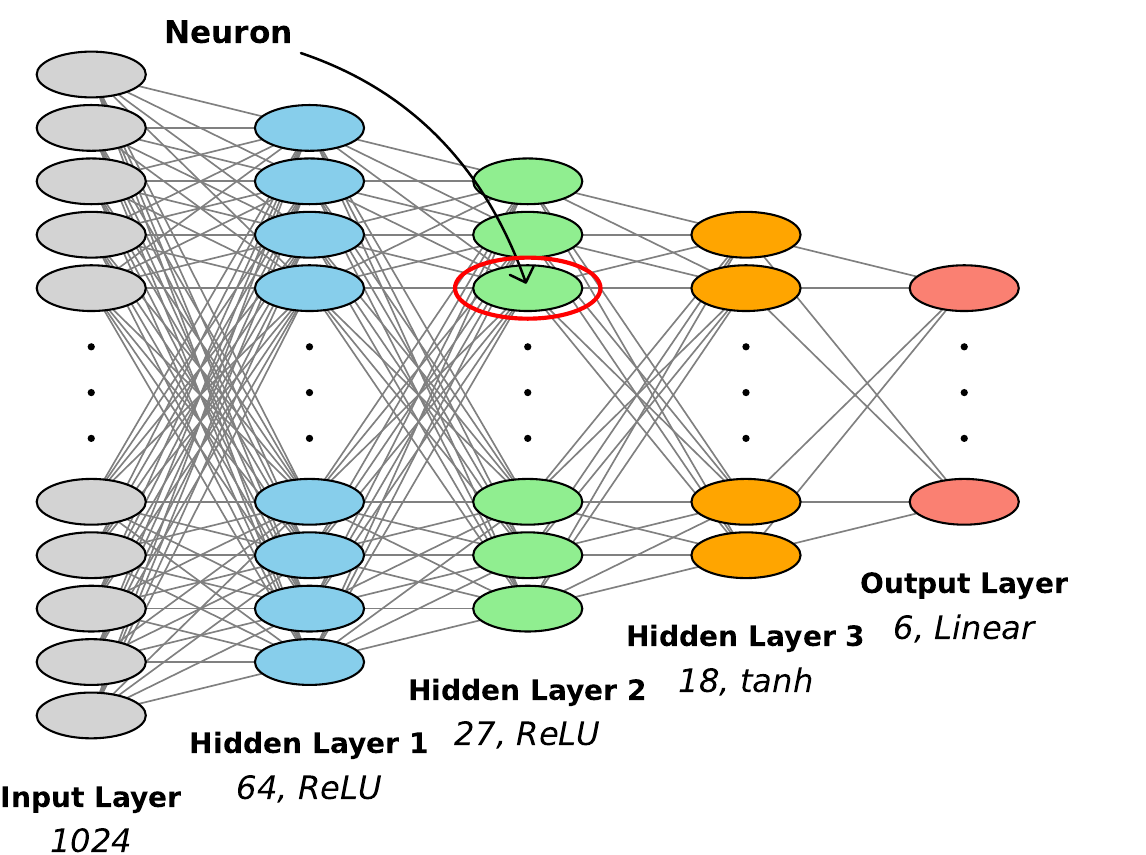}
    \caption{Architecture of the ANN used for parameter estimation. Each circle represents a neuron, which is fully connected to the neurons in the next layer. The first layer is the input layer, the final layer is the output layer, and the intermediate layers are hidden layers activated by their respective activation functions.}
    \label{fig:ann-architecture}
\end{figure}

To optimize the hyperparameters of the ANN, we used the \texttt{RandomSearchCV} package from \texttt{sklearn}. In the signal-only scenario, the neural network architecture consists of several fully connected layers. Thefirst hidden layer has 64 neurons, a ReLU activation function, and an input dimension of 1024. It is followed by a hidden layer with 27 neurons, also using the ReLU activation function. The next hidden layer contains 18 neurons, with a tanh activation function and weights initialized using a uniform distribution. The output layer consists of 6 neurons. A schematic of the ANN architecture is shown in Figure~\ref{fig:ann-architecture}. In the signal-with-foreground scenario, the architecture remains the same, with the exception of the output layer, which now contains 10 neurons. For the PCA preprocessing case, the input layer size is reduced to 100, while the rest of the architecture remains unchanged. 

\subsection{Support Vector Regression (SVR)}

Support Vector Regression (SVR) \citep{cortes1995support, vapnik1999overview} is a machine learning algorithm designed for regression tasks, with the goal of finding a function that approximates the relationship between input variables and a continuous output. SVR works by mapping the input data into a higher-dimensional space using a kernel function, and then identifying a hyperplane that maximizes the margin while ensuring the prediction error remains within a specified threshold. This approach is particularly effective for handling non-linear relationships and is known for its robustness against overfitting, especially in high-dimensional spaces. SVR seeks to determine the functional form of the output by finding a set of hyperplanes that effectively separate the training data.
SVR is trained in a way that the error between functional prediction $f(x_{i}) = w_{i}x_{i}$  for input data point $x_{i}$ and the actual output data point, $y_{i}$, lies within a margin , $\epsilon$, i.e.
\begin{align}
-\epsilon \le f(x_{i}) -y_{i} \le \epsilon
\label{constraint}
\end{align}
A relaxation is given to equation \ref{constraint}, in form of $(\xi, \xi^{\ast})$, such that,
\begin{align}
-\epsilon -  \xi^{\ast} \le f(x_{i}) -y_{i} \le \epsilon + \xi
\end{align}
The training of SVR tends to minimize the objective function,
\begin{align}
Objective = \frac{1}{2} |\mathbf{w}|^{2} + C\sum_{i = 1}^{n}|\xi_{i}|
\end{align}
Here, C is called the penalty term. For a non-linear space, kernel function takes it to a linear space so that $y_{i} = w_{i}x_{i}$. 
In dual form, the cost function is:
where, $k(\mathbf{x_{i},x_{j}})$, is the kernel function. Cost function is optimized in a way, so that:
\begin{align}
\sum_{i= 1}^{N}(\alpha_{i} - \alpha_{i}^{\ast}) = 0, \\ 
0 \le \alpha_{i},\alpha_{i} \le C
\end{align}
The kernel function used in our work is radial basis function (RBF) given as:
\begin{align}
k(\mathbf{x_{i},x_{j}}) = \exp(-\gamma|\mathbf{x_{i}} -\mathbf{x_{j}}|^{2})
\label{kernal}
\end{align}
The tunable hyperparameters are penalty $C$, $\gamma$ in the kernel equation \ref{kernal} and margin, $\epsilon$. To tune the hyperparameter of the SVR  we utilized the \texttt{RandomSearchCV} package from \texttt{sklearn}. For the signal case, the best hyperparameter for the SVR model is configured with a radial basis function (RBF) kernel, a polynomial degree of 10, a gamma value set to "auto," a coefficient coef0=2, a regularization parameter C=10, and an epsilon value of 0.01. Similarly, for the Signal with Foreground scenarios, the best performance with SVR is achieved when training each parameter individually with different hyperparameter combinations. The detailed architecture of the hyperparameters list is provided in the Table~\ref{tab:svr_hyperparameters}.

\begin{table}[h]
    \centering
    \begin{tabular}{|c|c|c|c|c|c|}
        \hline
    &$\gamma$ & $\epsilon$ & Degree & $\text{coef}_0$ & $C$ \\ 
        \hline
      Parm1  &  scale & 0.001 & 8 & 1.67 & 3.59 \\ \hline
      Parm2  &  scale & 0.001 & 8 & 1.67 & 3.59 \\ \hline
      Parm3  & scale & 0.001 & 8 & 1.67 & 3.59 \\ \hline
      Parm4  & auto  & 0.060 & 10 & 3.89 & 3.59 \\ \hline
      Parm5  & scale & 0.001 & 8 & 1.67 & 3.59 \\ \hline
      Parm6  & scale & 0.167 & 8 & 0.00 & 46.42 \\ \hline
      Parm7  & auto  & 0.003 & 9 & 2.22 & $10^{4}$ \\ \hline
      Parm8  & scale & 0.003 & 7 & 5.00 & 46.42 \\ \hline
      Parm9  & scale & 0.003 & 7 & 5.00 & 46.42 \\ \hline
      Parm10 & auto  & 0.060 & 10 & 3.89 & 3.59 \\ 
        \hline
    \end{tabular}
    \caption{Hyperparameter configurations for SVR for the signal with Foreground scenario.}
    \label{tab:svr_hyperparameters}
\end{table}


\subsection{Gaussian Process Regression (GPR)}
Gaussian Processes (GPs) are a flexible, non-parametric machine learning approach used for regression and classification tasks \citep{rasmussen2003gaussian}. They model data by assuming it is drawn from a multivariate Gaussian distribution, where the covariance is defined by a kernel function. GPs excel at capturing uncertainty and making predictions with confidence intervals, making them particularly useful in scenarios with limited data or complex relationships. Their performance depends heavily on the choice of kernel, which encodes prior assumptions about the data, such as smoothness or periodicity.
GPR is a non-parametric regression model which calculates the probability distribution, $P$, over all fitting functions, $\textbf{f}$, for the given data. A Gaussian process prior is assumed with a mean function, $\nu$, and covariance function, $K$. 
\begin{align}
log P(f)= -\frac{1}{2}\sum_{i,j = 1}{D}(f_{i} - \nu_{i})K_{ij}(f_{j} - \nu_{j}) + constant
\end{align}
Extending to multi variate Gaussian distribution to infinite dimensions, $\nu$ and $K$ assumes functional forms, $\nu(\mathbf{x})$ and $K(\mathbf{x})$. The prior for the covariance function is called the kernel. It is a tunable hyperparameter. 

The kernel used in our work is a MATERN kernel, which is a modified version of RBF kernel. We used SCIKIT LEARN to implement the GPR model in our work. To tune the hyperparameters for Gaussian Process Regression (GPR), we utilized the \texttt{RandomSearchCV} package from \texttt{sklearn}. For the signal-only case, the best combination of hyperparameters we found included a squared exponential (constant) kernel with a magnitude of 0.942, a Matérn kernel with a length scale of 24.8 and a smoothness parameter $\nu$=2.5, and a WhiteKernel that accounts for observational noise with a noise level of 0.00388. For the signal-plus-foreground scenario, the optimal hyperparameter combination consisted of a squared exponential (constant) kernel with a magnitude of 0.7072, a Matérn kernel with a length scale of 2 and smoothness parameter $\nu$=2.5, and a WhiteKernel modeling observational noise with a noise level of 0.5.

\subsection{Random Forest Regression (RFR)}
RFR is a regression method based on decision trees. There are N number of decision trees \citep{Breiman2001}. Each decision tree is trained on a randomly sampled set of datasets. The sampling is done with a replacement known as bootstrapping. The average of the predictions made by all the decision trees is returned as the output.

\begin{equation}
    Y_i = \frac{1}{N}\sum_{t=1}^{N} f_{t}(x)
\end{equation}

To determine the optimal architecture for the Random Forest Regressor (RFR), we utilized the \texttt{RandomSearchCV} package from \texttt{sklearn} to tune the hyperparameters and identify the best combination. For the signal-only case, the optimal hyperparameters for the random forest model are set as follows: nestimators=180, min samples split=7, min samples leaf=3, max features=’sqrt’, max depth=17, and bootstrap=False. Similarly, for the signal-plus-foreground scenario, the best hyperparameter combination includes nestimators=876, min samples split=3, min samples leaf=2, max features=’sqrt’, max depth=69, and bootstrap=True.


\section{Datasets preparation for Training and Testing}\label{ttd}
\begin{figure}
    \centering
    {\includegraphics[width=0.45\textwidth]{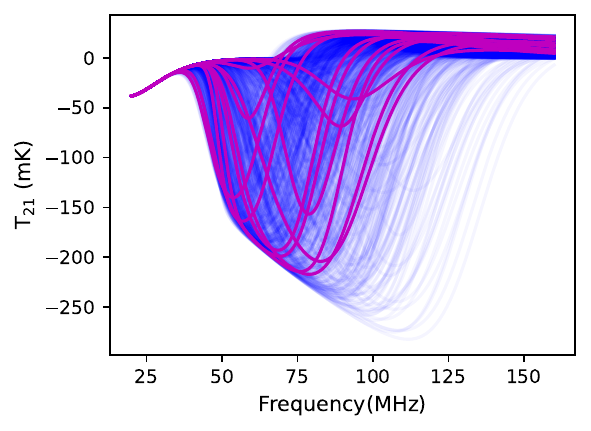}}
    \caption{Training dataset for the global 21-cm signal. The signal subsets are highlighted in red, while the remaining samples are shown in blue as background.}
   
    \label{Fig_sig_training}
\end{figure}

In this study, we created the training and testing datasets by sampling the parameter space using the Hammersley Sequence Sampling (HSS) \citep{HSS} method, as outlined by \cite{Tripathi2024_Samp}. In the signal-only case, we generated five datasets with sample sizes from 1,000 to 10,000 in increments of 2,500 to evaluate the performance of machine learning algorithms as the training dataset size increased. The datasets were produced using a parameterized ARES simulator, as detailed in Section \ref{ares}. The mock signal gallery presented in Figure~\ref{Fig_sig_training} was generated using Hammersley Sequence Sampling across the parameter ranges listed in Table~\ref{tab:1}. To train the machine learning models, we divided the datasets in a 70:30 ratio, allocating 70\% for training and 30\% for testing. 

\begin{figure}
    \centering
    {\includegraphics[width=0.45\textwidth]{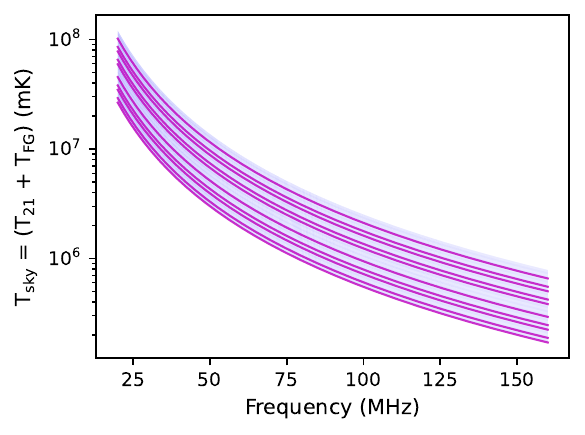}}
    \caption{Training datasets for the global 21-cm signal, including added foregrounds and thermal noise. A subset of the datasets is highlighted in red, while the remaining samples are shown in blue as background.}
   
    \label{Fig_sig_fore_training}
\end{figure}

In line with the signal-only case, for the scenario that included both foregrounds and thermal noise, we constructed five datasets with sample sizes of 10,000, 20,000, 30,000, 40,000, and 50,000. These datasets were designed to systematically evaluate how the performance of the machine learning model varies with increasing dataset size. These datasets were generated by sampling the parameter space using the Hammersley Sequence Sampling (HSS) method, based on the specified ranges of signal and foreground parameters outlined in Table \ref{tab:1}. A sample of the training datasets, including the signal with added foregrounds and thermal noise, is shown in Figure~\ref{Fig_sig_fore_training}.
Given the large dataset size, the data was split into a 9:1 ratio, with 90 \% allocated for training and 10 \% for testing the ML models. 

\begin{table}
\centering

\begin{tabular}{|c|c|c|}
\hline
& Parameters  & Ranges \\ \hline
& $\rm J_{z0}$  & 9.27, 27.81    \\
& $\rm X_{z0}$ & 4.34, 13.02    \\   
& $\rm T_{z0}$ & 4.89, 14.65     \\
Signal & $\rm J_{dz}$ & 1.65, 4.96  \\
& $\rm T_{dz}$ & 1.41, 4.23       \\
& $\rm X_{dz}$ & 1.42, 4.25     \\ \hline
& $\rm a_{0}$    &2.97, 3.64     \\
Foreground & $\rm a_{1}$  &-2.45,  -2.37 \\
&$\rm a_{2}$   & -0.082, -0.079  \\
&$\rm a_{3}$    & 0.027, 0.030  \\ \hline
\end{tabular}

\caption{The range of parameters used to build the training dataset for of global 21cm signals and foreground.}
\label{tab:1}
\end{table}

\section{Results and Discussions}\label{result}
In this study, we used four different machine learning regression models, including Gaussian process regression (GPR), random forest regression (RFR), support vector regression (SVR) and artificial neural networks (ANN), to extract global 21-cm signal parameters. This analysis is conducted in two scenarios: first, extracting parameters directly from the mock 21cm signal, and second, extracting both signal and foreground parameters from the mock signal embedded within the foreground and thermal noise. The performance of each ML model is evaluated based on parameter prediction accuracy using the $\rm R^{2}$ score and RMSE. Additionally, we assessed the computational efficiency of each model by measuring elapsed time and memory usage.     

\subsection{Signal only}
We trained each ML model using the global 21cm signal training dataset. Before training, the data set was preprocessed by normalizing and standardizing the input data with the StandardScaler function, while the corresponding parameters were normalized using the MinMaxScaler function, both from Scikit-learn. We used the same datasets to train and test each ML regression model to ensure consistency. Hyperparameter tuning for each model is performed using Scikit-learn's 'RandomizedSearchCV' API, which efficiently finds near-optimal hyperparameters by randomly sampling and evaluating their combinations. The final architecture is determined through this process. Each model is trained using the training data set and tested on a separate test dataset unknown to the trained network. The performance of the models is evaluated by calculating the $\rm R^{2}$ score and the RMSE. We compared the performance of theseML models using a limited training dataset of 1,000 samples. In this scenario, GPR and SVR outperformed ANN and RFR. GPR and SVR achieved higher accuracy in predicting the global 21cm signal parameters, with overall $\rm R^{2}$ scores of 0.76 and 0.75 and RMSE scores of 0.14, respectively. ANN and RFR showed lower accuracy, with $\rm R^{2}$ scores of 0.69 and 0.57, and RMSE scores of 0.16 for ANN and 0.19 for RFR. Among all models, RFR exhibited the poorest performance. 
The possible reason behind this behavior is that in low-dimensional cases with limited training data, SVR, and GPR generally outperforms ANNs because they incorporate strong smoothness priors through kernel functions, adapt their complexity to the dataset size, and naturally enforce regularization or uncertainty handling. This makes them well suited for modeling the relatively smooth dependence of the global 21cm signal on astrophysical parameters. In contrast, ANNs are high-capacity models that require larger datasets to exploit their flexibility; with few samples, they tend to overfit or fail to generalize effectively. RFR performs the worst in this regime because its piecewise-constant structure cannot capture the spectral trends of the global 21cm signal. With small datasets, random splits across trees further fragment the data, reducing the model’s ability to capture global correlations. Moreover, since RFR cannot extrapolate beyond the training distribution and tends to favor high-variance features, it is less effective at recovering the smooth, continuous relationships that characterize the astrophysical parameter space. 

\begin{figure*}
    \centering
    {\includegraphics[width=1.0\textwidth]{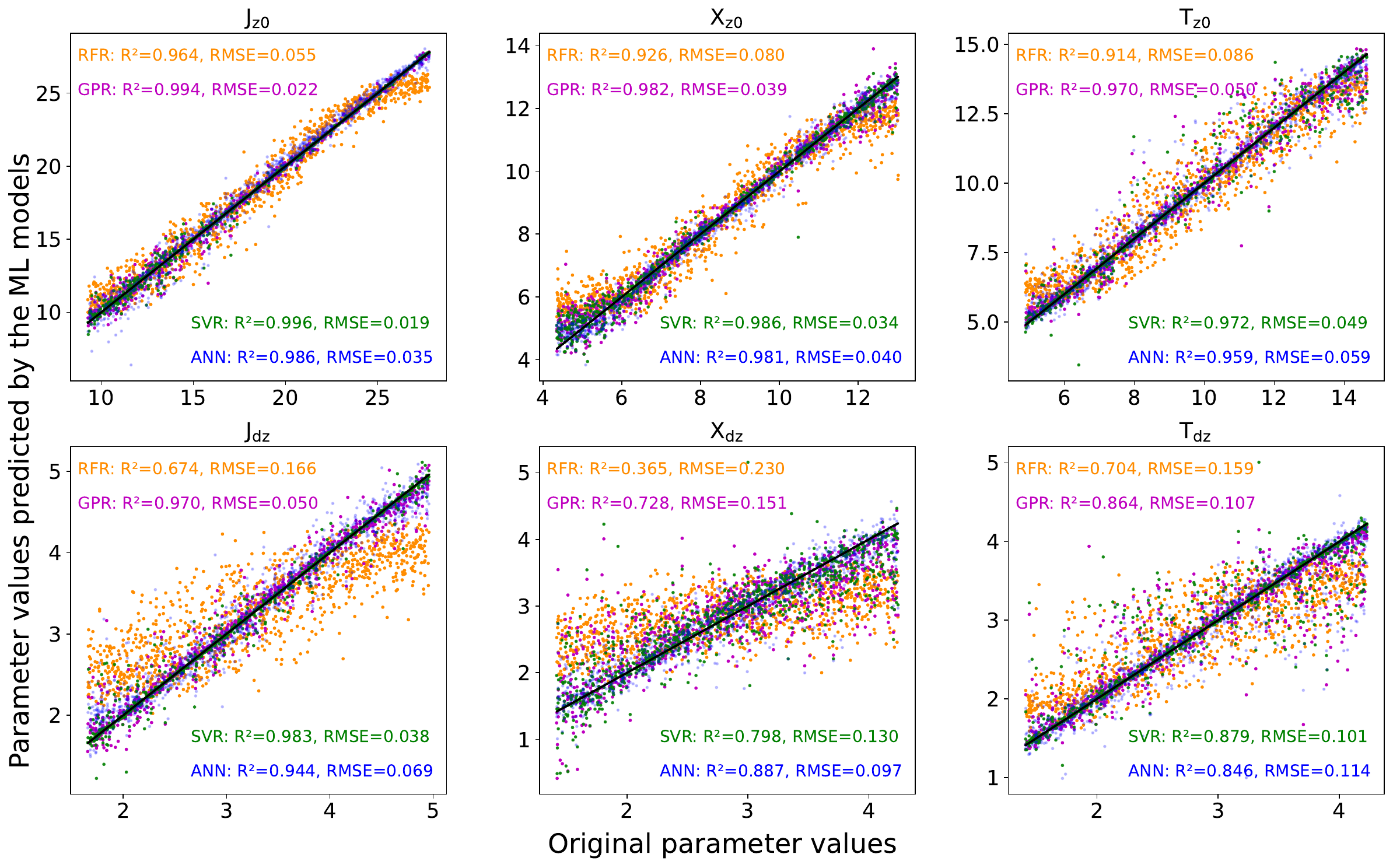}}
    \caption{The scatter plots show the predicted signal parameter values obtained using different machine learning models (RFR, GPR, SVR, ANN) trained on 10,000 signal-only datasets. In each plot, blue points represent ANN predictions, magenta points correspond to GPR, green points to SVR, and orange points to RFR. The solid black line represents the true parameter values, serving as a reference for model accuracy.}
   
    \label{Fig3}
\end{figure*}

\begin{figure*}
    \centering
    {\includegraphics[width=1\textwidth]{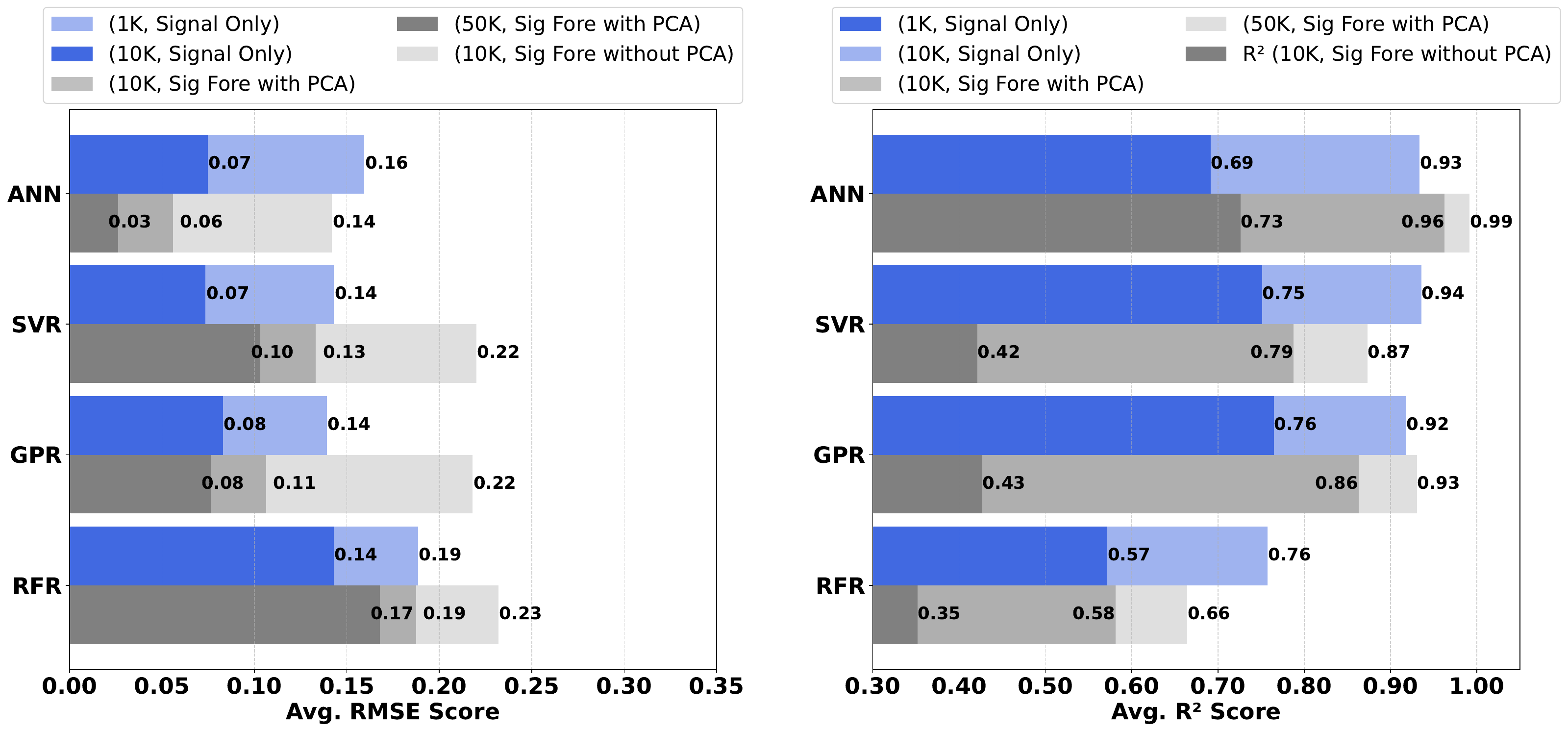}}
    \caption{Comparison of average RMSE and $\rm R^{2}$ scores across different machine learning models (RFR, GPR, SVR, ANN) for various dataset configurations. The left panel displays the RMSE scores, while the right panel presents the $\rm R^{2}$ scores. The models are evaluated on datasets of different sizes: signal-only datasets (1000, 10000) and signal-with-foreground datasets (10000, 50000). Additionally, the performance of each model is compared when trained on raw signal-with-foreground data versus data processed with PCA.}
   
    \label{Fig4}
\end{figure*}

\begin{figure}[!h]
    \centering
    {\includegraphics[width=0.48\textwidth]{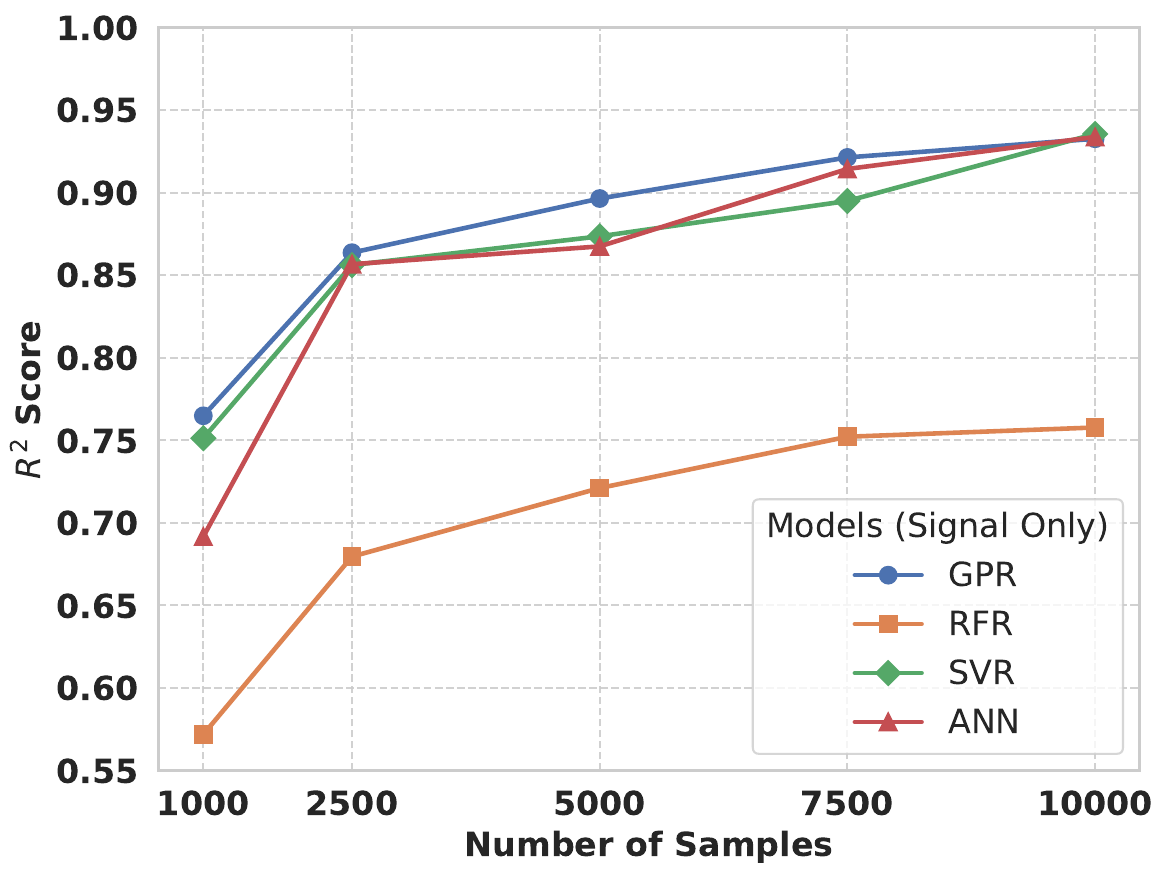}}
    \caption{Performance of different models for signal-only prediction as a function of training sample size. The $\rm R^{2}$ score increases with more samples, with GPR, SVR, and ANN achieving similarly high accuracy, while RFR consistently underperforms relative to the others.}
   
    \label{Fig7}
\end{figure}

To further assess the performance of the models, we increased the training dataset size to 10,000 samples. We visualized the prediction accuracy of each model for individual parameters using scatter plots, displaying both the $\rm R^{2}$ and RMSE scores (see Figure~\ref{Fig3}).
It led to a noticeable improvement in prediction accuracy across all models, with gains of approximately 18–20\%, as illustrated in Figure~\ref{Fig4}. Under this expanded dataset, the overall prediction accuracy of GPR, SVR, and ANN became comparable, although ANN maintained a slight edge over the others.
Among the models, ANN consistently delivered the most accurate predictions, achieving $\rm R^{2}$ scores between 0.986 and 0.846 and RMSE values in the range of 0.07–0.08 across all parameters. In contrast, both GPR and SVR struggled with the parameter $\rm X_{dz}$, yielding lower $\rm R^{2}$ scores of 0.728 and 0.798, respectively. Random Forest Regression (RFR) continued to show the least accurate performance among the four models. To systematically examine the impact of training dataset size on ML model performance, we varied the training set size from 1,000 to 10,000 in increments of 2,500, thereby also capturing performance variations at intermediate stages. Each ML model was trained on these datasets, and their performance was quantified using the $\rm R^{2}$ statistic. The results were then presented by plotting $\rm R^{2}$ as a function of dataset size (see Figure~\ref{Fig7}). 

We also compared the computational cost of running individual ML models in terms of memory usage and runtime; see details in Table~\ref{tab:computational_resource}. Among all four models, GPR required significantly more memory to process the 10,000 datasets, utilizing 4466 MB. In contrast, the other models were 3 to 4 times more memory-efficient for the same task. This indicates that GPR is more memory-intensive, while the other models are relatively efficient in memory usage. Regarding computational time, we observed that RFR was the fastest to execute but had the lowest prediction accuracy. Both ANN and GPR demonstrated moderate runtimes, whereas SVR required the longest runtime compared to the others. 

\begin{table*}
\begin{tabular}{|c|c|c|c|c|c|c|c|c|}
\hline

\textbf{\begin{tabular}[c]{@{}c@{}}Computational \\ Metric\end{tabular}} & \textbf{RFR} & \textbf{GPR} & \textbf{SVR} & \textbf{ANN} & \textbf{RFR} & \textbf{GPR} & \textbf{SVR} & \textbf{ANN} \\ \hline
& \multicolumn{4}{|c|}{\textbf{Signal Only 1000 Datasets}} & \multicolumn{4}{|c|}{\textbf{Signal Only 10,000 Datasets}} \\ \hline
\begin{tabular}[c]{@{}c@{}}Average RAM \\ Usage (GB)\end{tabular} & 0.51 & 0.80 & 0.55 & 0.64 &  1.62 & 4.36  & 1.19  & 1.60   \\ \hline
\begin{tabular}[c]{@{}c@{}}Elapsed Time  \\ (sec)\end{tabular} & 0.50 sec & 78sec & 2.60 sec & 34.3 sec  & 12.07 & 258.23 & 1302.22   & 236.51                 \\ \hline

& \multicolumn{4}{|c|}{\textbf{Signal with Foreground 10,000 Datasets}} & \multicolumn{4}{|c|}{\textbf{Signal with Foreground 50,000 Datasets}} \\ \hline
\begin{tabular}[c]{@{}c@{}}Average RAM \\ Usage (GB)\end{tabular} & 2.40 & 1.71  & 2.63  & 2.73 & 6.0 & 250 & 10.0 & 8.0 \\ \hline
\begin{tabular}[c]{@{}c@{}}Elapsed Time  \\ \end{tabular}  & 2.76 (sec) & 822.03 (sec) & 384.62 (sec) & 242.13(sec) &11.33 sec & 6 hr & 48 hr & 400 sec                \\ \hline

\end{tabular}

\caption{Computational performance metrics for different models for the same size of the  dataset.}
\label{tab:computational_resource}
\end{table*}

\subsection{Signal with Foreground and Thermal Noise}
We trained each ML model using a training dataset consisting of signal and foreground components with added thermal noise. Before training, we pre-processed the data using two different methodologies. In the first method, we applied logarithmic scaling to the dataset, followed by normalization and standardization, as described in \citep{Tripathi2024_Samp}. In the second method, we employed Principal Component Analysis (PCA) for dimensionality reduction. Specifically, the data dimensionality was reduced from 1024 to 100 features. All ML models were then trained using the data sets processed using this PCA-based approach. This appraoch we using first time for the global 21cm signal parameter extraction, earlier it used to make the emulator for the global 21cm by \cite{cohen2020emulating}. 
To tune the hyperparameters of the ML models, we followed a methodology similar to the one used for the signal-only case, utilizing RandomizedSearchCV to identify the optimal architecture for each model. The final architectures were trained and tested on 10,000 datasets, and the model prediction accuracy was evaluated using the $\rm R^{2}$ score and RMSE. In the first scenario, the trained models achieved the following results:GPR with $\rm R^{2}$ = 0.427 and RMSE = 0.218, RFR with $\rm R^{2}$ = 0.352 and RMSE = 0.232, SVR with $\rm R^{2}$ = 0.421 and RMSE = 0.220, and ANN with $\rm R^{2}$ = 0.726 and RMSE = 0.142. We computed the computational cost for each model, where GPR had an average memory usage of 2016.22 MB with an elapsed time of 1797.91 seconds, RFR required 1666.98 MB and completed in 7.83 seconds, SVR consumed 1625.66 MB while taking 6902.60 seconds, and ANN utilized 1198.49 MB with a runtime of 245.3 seconds.

\begin{figure*}
    \centering
    {\includegraphics[width=1.0\textwidth]{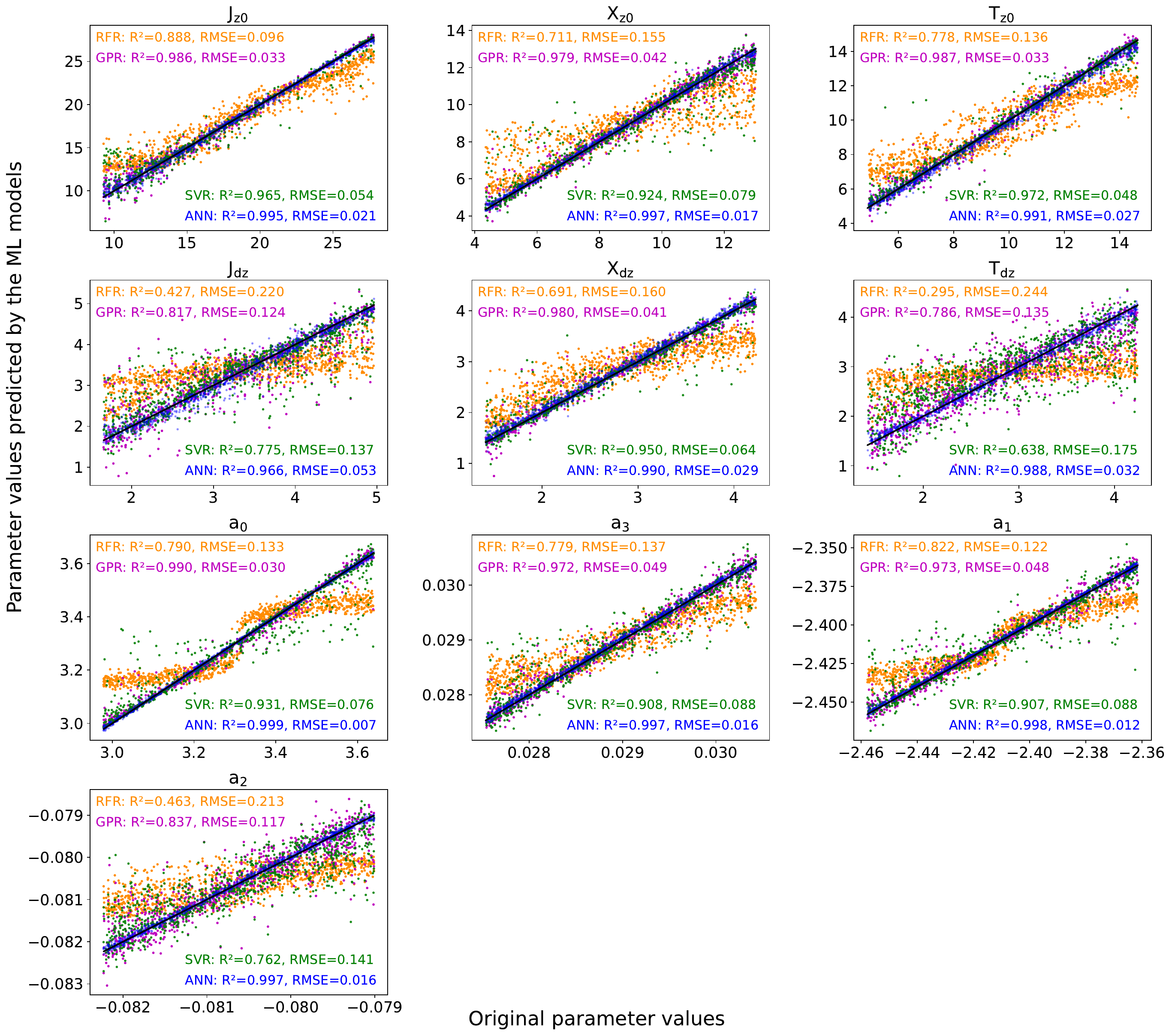}}
    \caption{The scatter plots illustrate the predicted values of signal and foreground parameters obtained using different machine learning models (RFR, GPR, SVR, ANN) trained on 50,000 datasets incorporating signal, foreground, and thermal noise. In each plot, blue points represent predictions from ANN, magenta points from GPR, green points from SVR, and orange points from RFR. The solid black line indicates the true parameter values, providing a reference for model accuracy.}
   
    \label{Fig6}
\end{figure*}

For the second scenario, when all the ML models were trained using PCA-processed datasets, we achieved improved accuracy with just 10,000 samples. The results were as follows: GPR with $\rm R^{2}$ score 0.862 and RMSE 0.106, RFR with $\rm R^{2}$ score 0.581 and RMSE 0.187, SVR with $\rm R^{2}$ score 0.756 and RMSE 0.133, and ANN with score $\rm R^{2}$ 0.958 and RMSE 0.056. Among the ML models, RFR continued to underperform compared to the other models, while SVR also struggled to accurately predict all parameters. In contrast, GPR and ANN delivered significantly better performance, with ANN achieving the highest accuracy. To assess the impact of larger training sets, we further evaluated the ML models using 50,000 samples; the corresponding results are shown in Figure~\ref{Fig6}, which shows the predicted parameter values plotted against the original values with different ML models. The figure also presents the corresponding $\rm R^{2}$ and RMSE scores for each model’s predictions of individual parameters. 

Training ML models with PCA-processed data sets significantly improves their accuracy from 20 \% to 40\%. Additionally, we computed the computational cost required to run each ML model on a dataset of 10,000 training samples. The GPR required an average memory of approximately 1752 MB and an elapsed time of 822.03 seconds. The RFR required around 2454.59 MB of memory, with an execution time of 2.76 seconds. The SVR utilizes an average memory of 2686.35 MB and takes 384.62 seconds to complete the task. Lastly, the ANN required approximately 2798.31 MB of memory, with an elapsed time of 242.13 seconds. Among the ML models, GPR used the least memory, while the other models required significantly more. From Table~\ref{tab:computational_resource}, we observed that the memory requirements of SVR, RFR, and ANN were comparable. In terms of execution time, RFR had the shortest runtime, whereas GPR required the longest time to complete.

In higher-dimensional cases with larger training datasets, ANNs outperform the other methods because they scale well with data and can capture complex, nonlinear relationships. Once smooth but strong foregrounds are added, the challenge shifts: the task becomes disentangling a small, non-smooth cosmological signal from a dominant, smooth contaminant. ANNs’ flexibility enables them to capture these subtle, non-smooth residual structures of the 21cm signal. In contrast, due to their smoothness priors, SVR tends to over-smooth, effectively fitting the foreground but washing out the cosmological signal. GPR remains somewhat competitive, as its kernel priors still represent smooth components effectively, but its computational cost limits scalability. In contrast, SVR performs poorly in this regime due to the curse of dimensionality, kernel inefficiencies, and the reduced effectiveness of distance-based similarity in high-dimensional spaces.

\begin{figure}[!h]
    \centering
    {\includegraphics[width=0.48\textwidth]{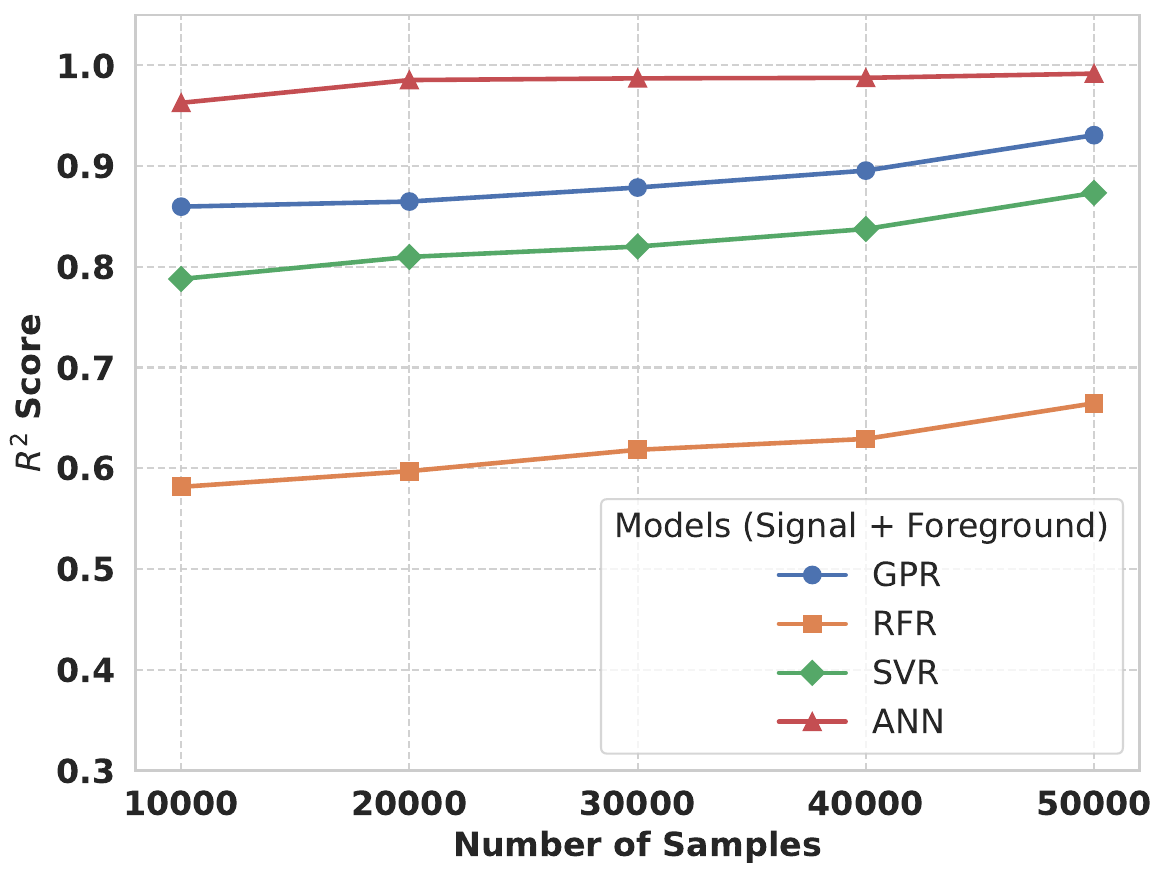}}
    \caption{Performance comparison of machine learning models for signal with added foreground and thermal noise prediction as a function of training sample size. The $\rm R^{2}$ score increases with the number of samples for all models, with the ANN consistently achieving the highest accuracy, followed by GPR, SVR, and RFR.}
   
    \label{Fig8}
\end{figure}

We also systematically examined the impact of training dataset size on ML model performance, following the same approach as in the signal-only case. The training set size was varied from 10,000 to 50,000 in increments of 10,000, thereby capturing performance variations at intermediate stages. Each ML model was trained on these datasets, and their performance was quantified using the $\rm R^{2}$ statistic. The results are presented in Figure~\ref{Fig8}, where $\rm R^{2}$ is shown as a function of dataset size. With a training set of 50,000 samples, we observed performance improvements of 4–10\% across all models: GPR achieved an $\rm R^{2}$ score of 0.931 with an RMSE of 0.0764, RFR obtained $\rm R^{2}=0.665$ with RMSE $=0.1679$, SVR reached $\rm R^{2}=0.8734$ with RMSE $=0.1032$, and ANN delivered the best performance with $\rm R^{2}=0.9918$ and RMSE $=0.0262$. However, the computational cost increased substantially, particularly for GPR, which required over 250 GB of memory and 6 hours to train, and SVR, which consumed 10 GB of memory but required 48 hours. In contrast, RFR and ANN were significantly more efficient, requiring 6 GB in 11.33 seconds and 8 GB in 400 seconds, respectively.



\section{Conclusion}\label{con}
In this work, we explore different machine learning regression models to enhance global 21cm signal parameter extraction, as well as foreground removal and parameter retrieval. The primary goal is to identify the most effective and efficient ML model for removing foreground contamination from the global 21cm signal while accurately extracting key astrophysical and intergalactic medium (IGM) parameters.

We compare the performance of four ML regression models: Random Forest Regressor (RFR), Gaussian Process Regressor (GPR), Support Vector Regressor (SVR), and Artificial Neural Networks (ANN). The performance of each model is evaluated using root mean square error (RMSE) and $\rm R^{2}$ scores.

Furthermore, we assess the models under different scenarios, including variations in training dataset size and dimensionality. In the signal-only case, the models are trained to predict six free parameters, whereas, in the presence of foregrounds and thermal noise, the models must predict ten free parameters. Additionally, we analyze the computational resources required for training each ML model across different dataset sizes and scenarios. The final results are summarized as follows:

\begin{itemize}
    \item For signal only case, GPR and SVR achieved the lowest RMSE and highest  $\rm R^{2}$ for small training datasets, outperforming ANN. Their performance further improved with more data, reaching RMSE values of approximately 0.0736–0.0829 and  $\rm R^{2}$ scores of 0.9182–0.9356 for 10,000 samples. ANN performed competitively, achieving an  $\rm R^{2}$ of 0.9338 for 10,000 samples. In contrast, RFR had the weakest performance, exhibiting the highest RMSE and lowest  $\rm R^{2}$ among all models.
    
    \item In the signal+foreground case (with PCA preprocessing), ANN significantly outperformed other models, achieving the lowest RMSE (0.0559 for 10,000 and 0.0262 for 50,000 samples) and the highest $\rm R^{2}$ ($\sim$ 0.9628–0.9918). GPR performed well but was less effective than ANN in handling foreground contamination. RFR and SVR struggled the most, exhibiting higher RMSE and lower $\rm R^{2}$ scores.

    \item PCA significantly improved performance, especially for ANN and GPR. Without PCA, all models suffered a decline in accuracy, with increased RMSE and decreased  $\rm R^{2}$ scores (e.g., ANN's RMSE rose from 0.0559 to 0.1420, and  $\rm R^{2}$ dropped from 0.9628 to 0.7260).

    \item Increasing the dataset size from 10,000 to 50,000 improved accuracy, with ANN achieving the best performance (RMSE = 0.0262,  $\rm R^{2}$ = 0.9918). GPR also benefited but incurred higher computational costs.

    \item \textbf{Memory Usage:} GPR had the highest RAM consumption (4.36 GB for 10,000 datasets, 250 GB for 50,000), making it impractical for large-scale use. ANN maintained moderate usage (8 GB for 50,000), while RFR was the most memory-efficient (1.62 GB for 10K).
    
    \item \textbf{Execution Time:} SVR and GPR were the slowest models, with SVR having the highest computational burden, taking 48 hours for 50,000 datasets, making it impractical for large-scale analysis. In contrast, ANN was much faster.
    
\end{itemize}

ANN emerges as the most effective model, offering a balanced trade-off between high accuracy, efficient memory usage, and reasonable computational cost, making it particularly well-suited for global 21 cm signal parameter extraction. A common limitation of regression-based methods is their inability to provide uncertainty estimates for the predicted parameters. In the case of ANNs, however, this can be mitigated by adopting a custom loss function, as demonstrated in \citet{2021Jeffrey} and \citet{2022Villaescusa}. In this work, we have employed such a custom loss function, with details provided in ~\ref{A1}; future studies will expand on this approach in greater depth. GPR, on the other hand, also delivers strong performance but comes with significant computational demands, including high memory requirements and long execution times. Its key advantage lies in its ability to naturally provide uncertainty estimates, a feature of considerable importance for cosmological and astrophysical applications. SVR, on the other hand, struggles with large datasets due to its high computational cost. PCA preprocessing plays a crucial role in enhancing model accuracy, especially in the presence of foreground contamination. While increasing the dataset size improves performance, computational efficiency remains a critical factor. In future work, we aim to extend this study by incorporating additional observational effects, such as ionospheric distortions and beam chromaticity, to make the analysis more applicable to realistic datasets. Furthermore, exploring other machine learning regression models could enhance both the accuracy and robustness of parameter extraction.


\appendix
\section{Custom Loss Function for Uncertainty Estimation in ANN}\label{A1}
To estimate the uncertainty in the ANN model, we adopt a custom loss function following the approach of \citet{2021Jeffrey, 2022Villaescusa}. In this framework, the network is trained to predict two quantities for each parameter: the mean ($\mu_i$) and the standard deviation ($\sigma_i$) of its marginal posterior distribution. These are defined as

\begin{equation}
\mu_i (X) = \int_{\theta_i}p(\theta_i \mid X)\theta_i d\theta_i,
\label{eq:mean}
\end{equation}

\begin{equation}
\sigma_i^2(X) = \int_{\theta_i} (\theta_i - \mu_i)^2p(\theta_i \mid X)d\theta_i ,
\label{eq:variance}
\end{equation}

where $p(\theta_i \mid X)$ denotes the marginal posterior for parameter $i$, obtained by integrating over all other parameters:

\begin{align}
p(\theta_i \mid X) &= \int_{\theta_i} p(\theta_1, \theta_2, \ldots, \theta_n \mid X) \,
d\theta_{1} \ldots d\theta_{i-1} \nonumber \\
&\quad d\theta_{i+1} \ldots d\theta_{n}. 
\label{eq:marginal}
\end{align}

Here, $X$ denotes a 2D map. Following the moments network approach presented in \citep{2021Jeffrey, 2022Villaescusa}, we can define the loss function such that the network output converges to the above quantities:

\begin{align}
\mathcal{L} &= 
\sum_{i=1}^{n} \log \left( 
    \sum_{j \in \text{batch}} (\theta_{i,j} - \mu_{i,j})^{2} 
\right) \nonumber \\
&\quad + 
\sum_{i=1}^{n} \log \left( 
    \sum_{j \in \text{batch}} \big( (\theta_{i,j} - \mu_{i,j})^{2} - \sigma_{i,j}^{2} \big)^{2} 
\right).
\label{eq:loss}
\end{align}

The ANN architecture was implemented for both the signal-only case and the case including foregrounds and thermal noise. The network consists of three hidden layers with 128, 54, and 36 neurons, respectively. For the signal-only case, the input layer contains 1024 neurons, corresponding to the input dimension, and the output layer consists of 12 neurons to predict the means and standard deviations of the parameters. All other hyperparameters are identical to those used in the baseline ANN architecture.

In the signal-plus-foreground scenario, the architecture is unchanged apart from the output layer, which is expanded to 20 neurons to account for the corresponding means and standard deviations. For the PCA-preprocessed case, the input dimension is reduced to 100, while the hidden and output layers remain the same.

\subsection{signal only}
In this case, the custom-loss ANN model predicts the signal parameters with an accuracy comparable to that of the standard ANN. Using 10,000 training datasets, the model achieves $R^{2}$ scores between 0.99 and 0.87 across different parameters, consistent with the performance of the traditional ANN approach. To assess robustness, we evaluated four randomly selected test scenarios, with the predicted parameters and their associated uncertainties presented in Table~\ref{tab:sig_ann}. For comparison, GPR was applied to the same test datasets, and the corresponding predictions and uncertainties are also reported in Table~\ref{tab:sig_ann}. In this setting, the custom-loss ANN yields substantially narrower uncertainty estimates than GPR. A plausible explanation is that GPR struggles with high-dimensional mappings and tends to overestimate uncertainties in regions of sparse training coverage, whereas the ANN leverages its representational capacity to constrain the parameter space more effectively.

\begin{table*}[ht]
\centering
\caption{Comparison of true and predicted signal parameters with $1\sigma$ uncertainties for selected sets.}
\label{tab:sig_ann}
\begin{tabular}{c|cccccc}
\hline
Set & J$_{\rm z0}$ & X$_{\rm z0}$ & T$_{\rm z0}$ & J$_{\rm dz}$ & X$_{\rm dz}$ & T$_{\rm dz}$ \\
\hline
 True & $17.085$ & $4.449$ & $7.281$ & $3.266$ & $2.828$ & $3.619$ \\
 ANN Pred & $16.445 \pm 0.337$ & $4.869 \pm 0.338$ & $7.320 \pm 0.193$ & $3.451 \pm 0.135$ & $2.463 \pm 0.377$ & $3.642 \pm 0.076$ \\
 GPR Pred & $17.054 \pm 1.213$ & $4.711 \pm 0.567$ & $7.253 \pm 0.639$ & $3.288 \pm 0.216$ & $2.898 \pm 0.185$ & $3.634 \pm 0.184$ \\
\hline
 True & $22.938$ & $4.992$ & $12.737$ & $2.312$ & $3.486$ & $2.152$ \\
 ANN Pred & $22.750 \pm 0.270$ & $5.114 \pm 0.342$ & $12.525 \pm 0.183$ & $2.208 \pm 0.118$ & $3.003 \pm 0.411$ & $2.171 \pm 0.066$ \\
 GPR Pred & $22.949 \pm 1.192$ & $5.580 \pm 0.558$ & $12.763 \pm 0.628$ & $2.305 \pm 0.213$ & $3.023 \pm 0.182$ & $2.139 \pm 0.181$ \\
\hline
 True & $16.085$ & $11.687$ & $12.834$ & $1.827$ & $3.626$ & $2.858$ \\
 ANN Pred & $14.452 \pm 1.242$ & $11.604 \pm 0.163$ & $13.688 \pm 1.189$ & $2.402 \pm 0.441$ & $3.625 \pm 0.151$ & $2.693 \pm 0.695$ \\
 GPR Pred & $14.031 \pm 1.359$ & $11.669 \pm 0.636$ & $14.694 \pm 0.716$ & $2.406 \pm 0.242$ & $3.599 \pm 0.207$ & $1.636 \pm 0.207$ \\
\hline
 True & $11.118$ & $11.699$ & $9.196$ & $2.817$ & $3.362$ & $4.063$ \\
 ANN Pred & $9.852 \pm 0.861$ & $11.703 \pm 0.179$ & $11.560 \pm 1.358$ & $2.539 \pm 0.378$ & $3.362 \pm 0.186$ & $3.090 \pm 0.674$ \\
 GPR Pred & $10.793 \pm 1.189$ & $11.856 \pm 0.556$ & $10.864 \pm 0.627$ & $2.720 \pm 0.212$ & $3.497 \pm 0.181$ & $3.030 \pm 0.181$ \\
\hline
\end{tabular}
\end{table*}

\subsection{Signal with Foreground and Thermal Noise}
Similar to the signal-only case, we trained the custom-loss ANN with signal with added foreground and thermal noise dataset. With 50,000 training samples, the model attained $\rm R^{2}$ scores in the range 0.99–0.95 across parameters, consistent with the performance of the traditional ANN model. Model robustness was further evaluated using four randomly selected test scenarios, with the predicted parameters and associated uncertainties reported in Table~\ref{tab1:sig_fore} for the signal parameters and Table~\ref{tab2:sig_fore} for the foreground parameters. For comparison, GPR was applied to the same test datasets, and the corresponding results are likewise presented in Table~\ref{tab1:sig_fore} and Table~\ref{tab2:sig_fore}. In this case as well, the custom-loss ANN provides substantially tighter uncertainty estimates than GPR, underscoring its effectiveness in constraining the parameter space.

\begin{table*}[ht]
\centering
\caption{Comparison of true and predicted astrophysical signal parameters with $1\sigma$ uncertainties for selected sets.}
\label{tab1:sig_fore}
\begin{tabular}{c|cccccc}
\hline
Set & J$_{\rm z0}$ & X$_{\rm z0}$ & T$_{\rm z0}$ & J$_{\rm dz}$ & X$_{\rm dz}$ & T$_{\rm dz}$ \\
\hline
 True & $25.582$ & $8.014$ & $12.704$ & $2.459$ & $1.804$ & $2.210$ \\
 ANN Pred & $25.523 \pm 0.116$ & $7.993 \pm 0.048$ & $12.752 \pm 0.130$ & $2.480 \pm 0.035$ & $1.789 \pm 0.052$ & $2.232 \pm 0.033$ \\
GPR Pred & $25.619 \pm 1.147$ & $8.240 \pm 0.537$ & $12.695 \pm 0.605$ & $2.427 \pm 0.205$ & $1.801 \pm 0.175$ & $2.283 \pm 0.175$ \\
\hline
True & $24.093$ & $7.532$ & $9.605$ & $4.303$ & $1.523$ & $2.171$ \\
ANN Pred & $23.987 \pm 0.185$ & $7.567 \pm 0.072$ & $9.644 \pm 0.154$ & $4.387 \pm 0.056$ & $1.517 \pm 0.043$ & $2.171 \pm 0.051$ \\
GPR Pred & $23.997 \pm 2.022$ & $7.644 \pm 0.946$ & $9.673 \pm 1.065$ & $4.326 \pm 0.361$ & $1.515 \pm 0.307$ & $2.135 \pm 0.309$ \\
\hline
True & $17.520$ & $10.929$ & $10.646$ & $4.125$ & $4.107$ & $1.457$ \\
ANN Pred & $17.315 \pm 0.293$ & $10.954 \pm 0.077$ & $10.734 \pm 0.174$ & $4.288 \pm 0.098$ & $4.094 \pm 0.058$ & $1.504 \pm 0.041$ \\
GPR Pred & $17.603 \pm 0.999$ & $11.248 \pm 0.468$ & $10.791 \pm 0.527$ & $4.086 \pm 0.178$ & $4.121 \pm 0.152$ & $1.730 \pm 0.153$ \\
\hline
 True & $13.672$ & $11.171$ & $11.307$ & $2.069$ & $1.691$ & $1.662$ \\
 ANN Pred & $13.616 \pm 0.375$ & $11.152 \pm 0.099$ & $11.724 \pm 0.210$ & $2.041 \pm 0.156$ & $1.672 \pm 0.060$ & $1.601 \pm 0.056$ \\
 GPR Pred & $13.532 \pm 2.657$ & $11.981 \pm 1.244$ & $11.810 \pm 1.400$ & $2.286 \pm 0.474$ & $1.611 \pm 0.404$ & $1.960 \pm 0.406$ \\
\hline
\end{tabular}
\end{table*}

\begin{table*}[ht]
\centering
\caption{Comparison of true and predicted foreground parameters with $1\sigma$ uncertainties for selected sets.}
\label{tab2:sig_fore}
\begin{tabular}{c|cccc}
\hline
Set & a$_{0}$ & a$_{3}$ & a$_{1}$ & a$_{2}$ \\
\hline
True & $3.14679$ & $0.02769$ & $-2.39634$ & $-0.07947$ \\
ANN Pred & $3.14263 \pm 0.00801$ & $0.02768 \pm 0.00003$ & $-2.39604 \pm 0.00107$ & $-0.07947 \pm 0.00006$ \\
GPR Pred & $3.14600 \pm 0.04095$ & $0.02776 \pm 0.00018$ & $-2.39642 \pm 0.00596$ & $-0.07971 \pm 0.00020$ \\
\hline
True & $3.42609$ & $0.03001$ & $-2.36837$ & $-0.08079$ \\
ANN Pred & $3.42515 \pm 0.00750$ & $0.03000 \pm 0.00003$ & $-2.36928 \pm 0.00082$ & $-0.08074 \pm 0.00007$ \\
GPR Pred & $3.42983 \pm 0.07215$ & $0.02998 \pm 0.00032$ & $-2.36061 \pm 0.01051$ & $-0.08036 \pm 0.00035$ \\
\hline
True & $3.37673$ & $0.02865$ & $-2.43038$ & $-0.08047$ \\
ANN Pred & $3.37815 \pm 0.00530$ & $0.02862 \pm 0.00002$ & $-2.43093 \pm 0.00073$ & $-0.08043 \pm 0.00004$ \\
GPR Pred & $3.37877 \pm 0.03566$ & $0.02859 \pm 0.00016$ & $-2.43205 \pm 0.00519$ & $-0.08068 \pm 0.00017$ \\
\hline
True & $3.27920$ & $0.02883$ & $-2.41672$ & $-0.08070$ \\
ANN Pred & $3.27782 \pm 0.00815$ & $0.02881 \pm 0.00002$ & $-2.41611 \pm 0.00102$ & $-0.08064 \pm 0.00005$ \\
 GPR Pred & $3.27204 \pm 0.09482$ & $0.02903 \pm 0.00042$ & $-2.41822 \pm 0.01381$ & $-0.08084 \pm 0.00046$ \\
\hline
\end{tabular}
\end{table*}

\section{Hyperparameter configurations for SVR}\label{B1}

\section*{Acknowledgements}
AT would like to thank the Indian Institute of Technology Indore for providing funding for this study in the form of a Teaching Assistantship. AT would like to thank Suman Majumdar for the insightful discussions and continuous support. GK is supported by the Polish National Science Center through grant no. 2020/38/E/ST9/00395. The author acknowledge the use of facilities procured through the funding via the Department of Science and Technology, Government of India sponsored DST-FIST grant no. SR/FST/PSII/2021/162 (C) awarded to the DAASE, IIT Indore.




\bibliography{ref_revised2}
\end{document}